\documentclass[aip,apl,reprint,twocolumn,groupedaddress]{revtex4-1}
\usepackage{xcolor}
\usepackage{hyperref}
\hypersetup{colorlinks=true, citecolor=blue, urlcolor=blue, linkcolor=blue}
\usepackage{graphicx}
\usepackage{epstopdf}
\usepackage{subfigure}
\usepackage{amsmath}
\usepackage[version=3]{mhchem}

\begin{document}

\title{A first-principles study on the lattice thermal conductivity of irradiated glassy states of the \ce{Ge2Sb2Te5} phase-change memory material}

\author{Felix C. Mocanu}
\email[]{fcm29@cam.ac.uk}
\affiliation{Department of Chemistry, University of Cambridge, Lensfield Road, CB2 1EW, Cambridge, United Kingdom}

\author{Konstantinos Konstantinou}
\affiliation{Department of Chemistry, University of Cambridge, Lensfield Road, CB2 1EW, Cambridge, United Kingdom}

\author{Stephen R. Elliott}
\affiliation{Department of Chemistry, University of Cambridge, Lensfield Road, CB2 1EW, Cambridge, United Kingdom}

\date{\today}

\begin{abstract}
An analysis of thermal transients from non-equilibrium \emph{ab initio} molecular-dynamics simulations can be used to calculate the thermal conductivity of materials with a short phonon mean-free path. We adapt the approach-to-equilibrium methodology to the three-dimensional case of a simulation that consists of a cubic core region at higher temperature approaching thermal equilibrium with a thermostatted boundary. This leads to estimates of the lattice thermal conductivity for the glassy state of the phase-change memory material, \ce{Ge2Sb2Te5}, which are close to previously reported experimental measurements. Self-atom irradiation of the material, modelled using thermal spikes and stochastic-boundary conditions, results in glassy models with a significant reduction of the lattice thermal conductivity compared to the pristine glassy structure. This approach may prove to be useful in technological applications, e.g. for the suppression of thermal cross-talk in phase-change memory and data-storage devices.
\end{abstract}

\maketitle


The cost of computation is significantly determined by the energy needed to keep key components cool, whether it is central processing units, accelerators, memory or data storage. Even when components are kept at reasonable temperatures during normal operation, there may still be thermal-throttling under peak-load which affects their performance and lifetime. Hence, the ability to predict the thermal properties of solids is critical for the rational design of materials and the management of devices that are sensitive to thermal fluctuations.~\cite{TCSolids1,TCSolids2}

Phase-change memory devices, which are widely believed to be a part of at least some commercial non-volatile solid-state drives, rely on the fast and reversible switching between a conductive crystalline phase (the ``1'' state of a bit), and a resistive glassy phase (the ``0'' state of a bit), of the same chalcogenide material, induced via Joule heating from the application of appropriate voltage pulses.~\cite{Elliott2015} Consequently, thermal cross-talk between adjacent memory cells can limit the size down-scaling in phase-change memory devices.~\cite{Fong2017} An understanding of the thermal-transport properties of these materials at the atomic level is therefore essential for their design and optimization.~\cite{Lencer2011}

The lattice thermal conductivity of the phase-change memory material, GeTe, has been predicted successfully with quasi-static calculations,~\cite{Sosso2012} as well as with classical equilibrium,~\cite{Campi2015} and non-equilibrium,~\cite{Campi2015, Sosso2018} molecular-dynamics methods by using a linear-scaling neural-network interatomic potential.~\cite{Sosso2012} First-principles calculations were also employed to estimate the thermal conductivity for the crystalline phases of different chalcogenide phase-change materials, including \ce{Ge2Sb2Te5},~\cite{Campi2017} in good agreement with experimental observations. However, there are no previously reported modelling studies of the lattice thermal conductivity for the glassy phase of \ce{Ge2Sb2Te5} which is of equal importance with the crystalline counterpart for technological applications.

Methods for simulating thermal transport at the atomic level have seen a rapid evolution and are getting closer to quantitative agreement with experimental measurements.~\cite{Baroni2018, Sosso2018} \emph{Ab initio} methods, based on the Boltzmann transport equation,~\cite{BTE1,BTE2} or on Green-Kubo dynamical  formulations,~\cite{GK1,GK2} have been employed recently in the literature to provide a first-principles description of thermal transport, and they represent significant advances in the field. Nevertheless, these approaches are computationally very demanding within a density-functional-theory framework, while some of them are specifically designed for harmonic solids near equilibrium.

Therefore, there is the necessity for an efficient first-principles molecular-dyanamics approach to model the thermal properties of glasses and in particular the glassy phase of \ce{Ge2Sb2Te5}. An efficient and quantitative assessment of lattice thermal conductivity can be obtained from the approach-to-equilibrium molecular-dynamics (AEMD) method,~\cite{Lampin2012, Lampin2013} which has been successfully deployed in \emph{ab initio} (as well as with empirical force fields) molecular-dynamics simulations of several different materials.~\cite{Puligheddu2017} 

In this Letter, the AEMD methodology, which belongs to a larger class of non-equilibrium molecular dynamics (NEMD) methods, has been adapted to the case of a cubic core region in contact with a thermostatted boundary shell in order to calculate the thermal conductivity of irradiated glassy \ce{Ge2Sb2Te5}. Below, we briefly describe the computational details of the NEMD simulation protocol using stochastic-boundary conditions that is used to simulate the energetic thermal spikes. Numerical results are presented, starting with the fitting of thermal transients, the predicted thermal conductivity and the effect of irradiation on the thermal properties. We also compare our findings to experimental data and previous simulation studies, and we discuss some of the limitations of the approach.

The simulated system is a 315-atom melt-quenched model of glassy \ce{Ge2Sb2Te5}.~\cite{Lee2017} The simulation box, which has a length of 21.65 \AA{}, was divided into a core cubic region and an outer boundary shell of thickness 1 \AA{} on each side. This type of separation has been dubbed stochastic-boundary conditions, and was originally used to investigate thermal transport at interfaces.~\cite{Tenenbaum} The core region samples a micro-canonical ensemble (NVE) while the boundary undergoes Langevin dynamics in the canonical ensemble (NVT) and dissipates the heat generated in the core region, during an ionic cascade.~\cite{Toton2010} The thermostat parameters were chosen such that the damping period was 100 fs. \emph{Ab initio} molecular-dynamics simulations were carried out using the CP2K code,~\cite{Vandevondele2005} in which the stochastic-boundary-conditions approach is implemented, based on the Generalized Langevin Equation formulation.~\cite{Kantorovich2008} We modelled radiation-induced non-equilibrium cascades by performing thermal-spike simulations with initial kinetic energies in the range of $15-200$ eV. Further details related to the computational set-up, the radiation-damage cascades and the \emph{ab initio} molecular-dynamics simulations can be found in our previous work,~\cite{Konstantinou2018, Konstantinou2018a} where some of the initial trajectories used in this work have been taken from.

In the approach-to-equilibrium methodology, heat conduction is usually modelled by Fourier's law.~\cite{Lampin2013} Based on our set-up, the simulation box is cubic and is assumed to be approximately homogeneous and isotropic. Additionally, we have considered that the boundary region acts as a ``thermal wall" at a temperature of $T_0$ ($300$ K) and the core region will rapidly come into equilibrium with it after the cascade. In practice, the temperature of the boundary region will oscillate significantly and there will be an artificial thermal boundary resistance at the interface with the core region due to the thermostat that is employed in the outer shell.~\cite{Kapitza1941, Tenenbaum, Singh2015a} The Cartesian coordinates of atoms in the core region, $x$, $y$ and $z$, reside in the real interval $[0, L]$ where $L = a - 2r$, $a$ is the size of the periodic cubic simulation box and $r$ is the thickness of the boundary region. Under these assumptions, the resulting heat equation can be written as:

\begin{equation}
\label{eq:heat}
\dfrac{\partial T}{\partial t} = -\alpha \left( \dfrac{\partial^2 T}{\partial x^2} +  \dfrac{\partial^2 T}{\partial y^2} + \dfrac{\partial^2 T}{\partial z^2}\right)
\end{equation}
where $T$ is the temperature and $t$ is the time. 

The thermal diffusivity $\alpha$ is defined as:

\begin{equation}
\label{eq:alpha}
\alpha = \dfrac{\kappa}{C_v \rho}
\end{equation}
where $\kappa$ is the thermal conductivity, $C_v$ the constant-volume heat capacity and $\rho$ the density of the system. The general solution, assuming a separable form, is then given by:
\begin{equation}
\begin{split}
T(x,y,z, t) = T_0 + \sum_{n=1}^{\infty}\sum_{m=1}^{\infty}\sum_{l=1}^{\infty} a_{nml} e^{-\alpha \lambda_{nml} t} \\ \sin\left( \dfrac{n \pi x}{L}\right)\sin\left( \dfrac{m \pi y}{L}\right) \sin\left( \dfrac{l \pi z}{L}\right)
\end{split}
\label{eq:fourier}
\end{equation}

The Fourier-series coefficients in the general solution, $\lambda_{nml}$ and $a_{nml}$, can be inferred from the boundary conditions and are given in the two equations below:

\begin{equation}
\label{eq:lambda}
\lambda_{nml} = \left(\dfrac{\pi}{L}\right)^2 \left( n^2 + m^2 + l^2 \right)\\
\end{equation}

\begin{equation}
\label{eq:acoeff}
\begin{split}
a_{nml} = -\Delta T_0 \left(\dfrac{2}{\pi}\right)^3 \dfrac{\left(1 - (-1)^n\right)\left(1 - (-1)^m\right)\left(1 - (-1)^l\right)}{nml}
\end{split}
\end{equation}

The difference between the spatially averaged temperature of the core region (undergoing NVE dynamics) and the target temperature of the thermostatted boundary region (undergoing Langevin NVT dynamics) corresponds to: $\Delta T(t) = \bar{T}(t) - T_0$. It has an initial value, $\Delta T_0$, at the start of the thermal quench and it decays to zero during the quench.

\begin{equation}
\label{eq:exp}
\Delta T  \sim e^{-\dfrac{t}{\tau}}
\end{equation}

The exponential temporal decay of this temperature difference in equation~\ref{eq:exp}, has a dominant contribution from the leading term  $n=m=l=1$ of the Fourier series in equation~\ref{eq:fourier}. Once the relaxation time $\tau$ is obtained from simulations, it can be inserted into the time-dependent part of this dominant term. Hence, the thermal conductivity, $\kappa$, can be calculated from the expression:

\begin{equation}
\label{eq:kappa}
\kappa = \dfrac{L^2}{3 \pi^2} \dfrac{C_v \rho}{\tau}
\end{equation}

\begin{figure}[!ht]
\includegraphics[width=\columnwidth]{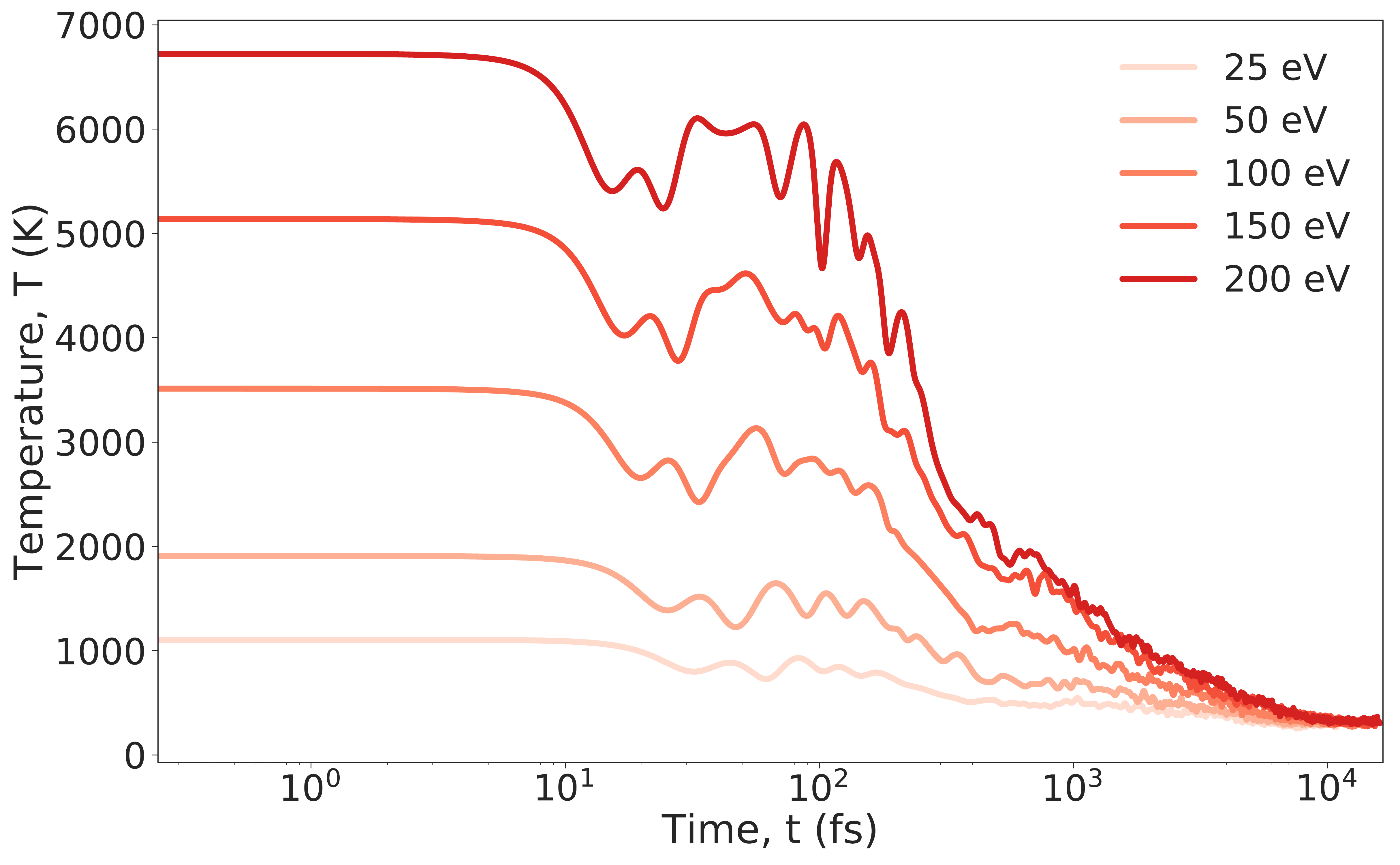}
\caption{Temperature (K) versus time (fs), on a linear-log graph, for a model of glassy \ce{Ge2Sb2Te5}, for different thermal-spike energies, shown as shades of red (darker means higher energy). The characteristic dips in temperature correspond to collision events during the non-equilibrium cascade. This is followed by an exponential decay of the temperature as the system approaches equilibrium, while at the end, the kinetic temperature of the system fluctuates around 300 K.}
\label{fig:logxTt}
\end{figure}

The thermal transients of the non-equilibrium ion-irradiation simulations for glassy \ce{Ge2Sb2Te5} can be directly fitted from the thermal quench of the core region as it reaches thermal equilibrium with the boundary layer. The trajectory of the radiation-induced cascade can be split into three intervals, based on the time evolution of the kinetic temperature, shown in figure~\ref{fig:logxTt} for different initial thermal-spike kinetic energies: (a) The high-energy cascade generated by the thermal spike; (b) An approach-to-equilbrium transient that is reasonably well described by an exponential temporal decay of the temperature difference between the core and the boundary; and (c) An equilibrium region where the system as a whole fluctuates around the target temperature of the thermostatted boundary layer (300 K).

\begin{figure}[htb]
\includegraphics[width=\columnwidth]{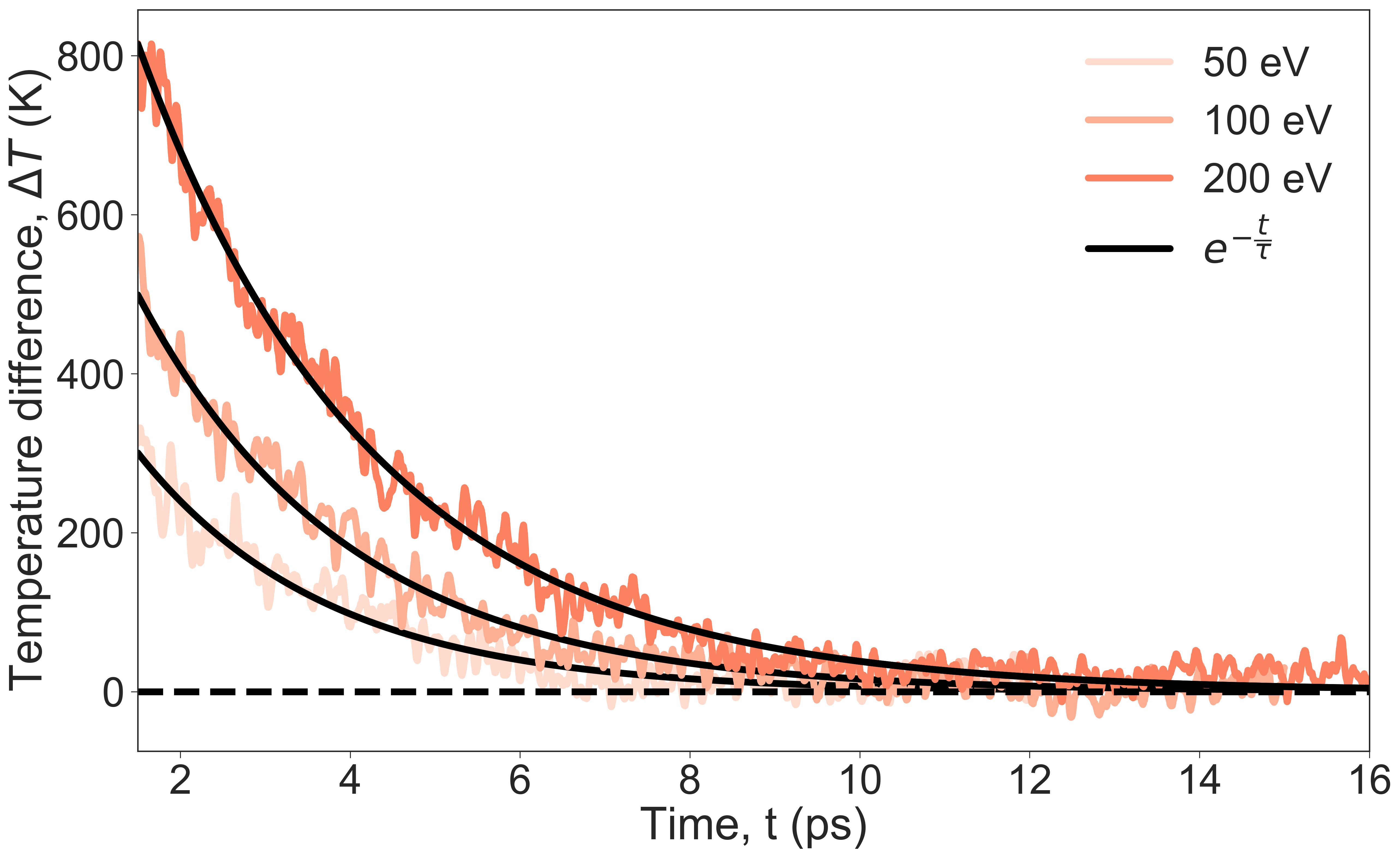}
\caption{The temperature difference $\Delta T$ between the core region and the target temperature of the boundary thermostat, viz. 300 K, as a function of time. Thermal transients in the approach-to-equilibrium are shown in different shades of red as a function of the energy of the thermal spike (darker is higher). The corresponding exponential fits are shown as black lines.}
\label{fig:expfit}
\end{figure}

By examining the approach-to-equilibrium of the system after the high-energy cascade, thermal-conductivity estimates are obtained ``on the fly'' without having to run a separate simulation for this purpose. It is important to only fit the relaxation time $\tau$ using data from a restricted time interval in which the temperature decays exponentially. In practice this means including data only after the shock of the thermal spike has been absorbed by the boundary and just before reaching equilibrium. In order to avoid including data from the high-energy cascade at the beginning of the simulation, a time period of $1-2$ ps from the start of the simulation needs to be removed from the fitting interval, based on the initial energy of the thermal spike, as indicated by our kinetic analysis of the approximate cascade duration.~\cite{Konstantinou2018a} Examples of exponential fits of the thermal relaxation time from the asymptotic regime of the ion-irradiation simulations are shown in figure~\ref{fig:expfit} for 50 eV, 100 eV and 200 eV initial thermal-spike energies.

After obtaining the transients for all the thermal-spike simulations, we examined the thermal conductivity of the initial pristine glassy \ce{Ge2Sb2Te5} structure. The computational procedure used in this case comprised the following steps: (1) Fix the atoms in the boundary region; (2) Initialize and equilibrate the velocities in the core region at a higher temperature; (3) Release the constraints for the boundary region; and (4) Remove the thermostat from the core region. In this case, there is no radiation-induced cascade and the simulation consists simply of a thermal quench and equilibration with the boundary. The core-region initial maximum kinetic temperature was chosen to be 700 K in order to avoid any intermixing between the core and the boundary.

\begin{figure}[htb]
\includegraphics[width=\columnwidth]{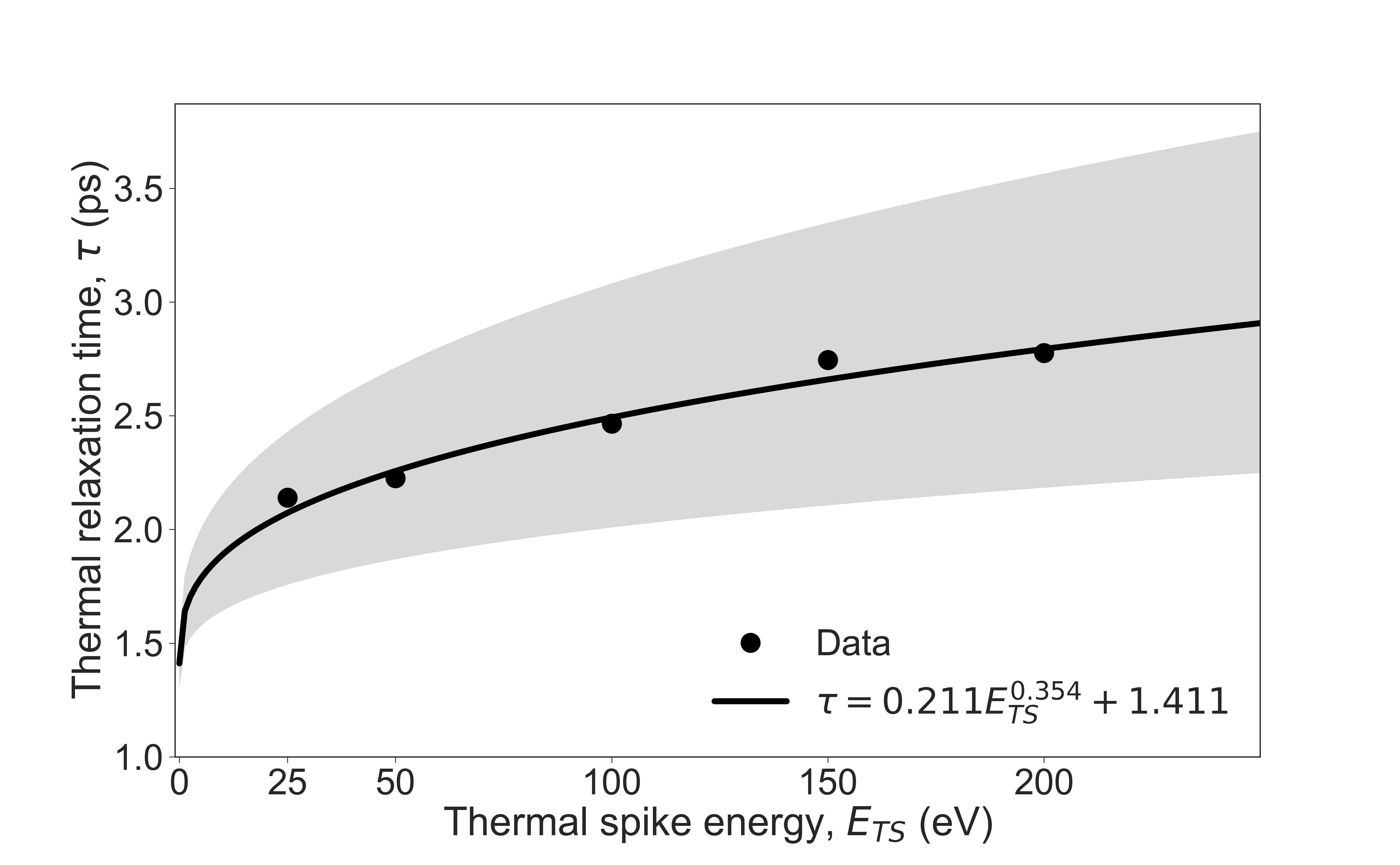}
\caption{ Thermal relaxation times versus the initial kinetic energy of the thermal-spike atom. Data points are shown as black circles. A power-law fit to the ion-irradiation data is shown as the solid black line and the shaded region is the 95\% confidence interval of the fit.}
\label{fig:powerlaw}
\end{figure}

The thermal relaxation times corresponding to the different ion-irradiation simulations and to the pristine glassy  \ce{Ge2Sb2Te5} structure were calculated, and are shown in figure~\ref{fig:powerlaw}. It can be observed that the relaxation time $\tau$ scales as a power-law function of the initial energy of the thermal spike. The temperature evolution is regularly analyzed after simulations of thermal spikes and a mechanical model predicts that the thermal relaxation time will scale as a power-law function of the thermal-spike energy, with an exponent around $2/3$.~\cite{Marks1997} However, values for the power-law exponent below $2/3$ have been reported from computer simulations, suggesting that the exponent depends on the structure of the material.~\cite{Buchan2015} From our simulations, a best-fit power-law exponent of $0.354$ was obtained for glassy \ce{Ge2Sb2Te5}, indicating a gentle increase of the thermal relaxation time with the energy of the thermal spike. We expect that this trend will not be significantly changed if the calculation is repeated for several independent amorphous models or indeed a larger model to accommodate higher thermal-spike energies. 

An additional ingredient for the estimation of the lattice thermal conductivity is the heat capacity of the material. In order to obtain the heat capacity, energy fluctuations, $\delta E = E - \langle E \rangle$, were sampled from equilibrium \emph{ab initio} molecular-dynamics simulations. The starting point for each trajectory corresponds to the structural models at the end of each non-equilibrium thermal-spike simulation. In these subsequent molecular-dynamics runs, there is no longer a separation between core and boundary, and the entire system undergoes NVT dynamics with a single Langevin equation (GLE), or colored-noise, thermostat.~\cite{Ceriotti2009, Ceriotti2010} Trajectories of 40 ps were generated with a time-step of 1 fs for each glassy sample. The first 8 ps were discarded and the heat capacity was estimated from the remaining 32000 time steps. 

The calculated heat capacity at 300 K for the pristine glassy phase of  \ce{Ge2Sb2Te5} was 272 $\pm$ 60.19 J kg$^{-1}$ K$^{-1}$, which is above the Dulong-Petit limit value of 219 J kg$^{-1} $K$^{-1}$ ( $3 N k_B$). This is in good agreement with experimental data from differential scanning calorimetry, viz. $220-230$ J kg$^{-1}$ K$^{-1}$,~\cite{Kuwahara2007, Zalden2014} and with the fact that the Debye temperature of glassy (as-deposited) \ce{Ge2Sb2Te5} was found to be below 111 K from inelastic neutron-scattering experiments.~\cite{Zalden2014} In the Supplemental Material, details are provided for the calculation of the heat capacity and thermal conductivity in all of the simulated systems.

The lattice thermal conductivity of glassy \ce{Ge2Sb2Te5} was calculated from equation~\ref{eq:kappa}, using a relaxation time corresponding to the intercept from our power-law extrapolation. This estimated thermal conductivity was found to be 0.16$\pm$0.04 W K$^{-1}$ m$^{-1}$ in good agreement with the experimentally reported values, which are in the range $0.19-0.3$ W K$^{-1}$ m$^{-1}$.~\cite{Peng1997,Giraud2005,Lyeo2006} From figure~\ref{fig:powerlaw}, it can be seen that the irradiated glassy \ce{Ge2Sb2Te5} models exhibit an increased thermal-relaxation time which ultimately leads to a decreased thermal conductivity. The estimated  lattice thermal conductivity of the self-irradiated configurations was found to be in the range of 0.067 to 0.113 W K$^{-1}$ m$^{-1}$, depending on the thermal-spike energy, further revealing that the thermal conductivity could be significantly lowered by controlled irradiation. It is noted that controlled irradiation (with \ce{He} ions) has been used successfully before to lower the lattice thermal conductivity of Si nanowires.~\cite{Zhao2017} 

In a recent AEMD study for a related glassy chalcogenide material, namely \ce{GeTe4}, the authors reported a value of 0.013$\pm$0.003 W K$^{-1} $m$^{-1}$ for the thermal conductivity,~\cite{Bouzid2017} which is an order of magnitude lower than the experimental measurement for this material (0.1 W K$^{-1}$ m$^{-1}$).~\cite{Zhang2009} This discrepancy is likely due to the small model system size (185 atoms), which illustrates the limitations of tractable \emph{ab initio} molecular-dynamics simulations. The same authors, in a different study,~\cite{Martin2018} doubled the simulated system size (up to 370 atoms) and obtained a value of 0.044$\pm$0.001 W K$^{-1}$ m$^{-1}$ for the thermal conductivity of glassy \ce{GeTe4}, closer to, but still smaller than, the experimental value, revealing the influence of size effects. 

The relatively low estimated value for the thermal conductivity of glassy \ce{Ge2Sb2Te5} in this work could suggest that the system size is still somewhat too small to fully account for the contribution of any long-wavelength vibrational modes to the lattice thermal conductivity.~\cite{Allen1993, Sosso2018} While not a substitute for analysing the finite-size effects, the influence of the boundary region Langevin-thermostat damping time was included in our estimates of the lattice thermal conductivity and is discussed in the Supplemental Material. 

In conclusion, a non-equilibrium molecular-dynamics methodology is proposed for calculating the lattice thermal conductivity of a cubic-core region as it approaches equilibrium with a thermostatted-boundary layer. This approach has been applied to estimate the change in the lattice thermal conductivity with thermal-spike energy in self-irradiated glassy \ce{Ge2Sb2Te5} models. As the irradiated core approached equilibrium with the thermostatted boundary, the thermal relaxation time is fitted using an appropriate time interval from the asymptotic regime of the simulation. The good agreement obtained between the calculated value for the lattice thermal conductivity of pristine glassy \ce{Ge2Sb2Te5} and the results of experimental measurements provides validation for this approach. 

The thermal relaxation time is found to increase as a sublinear power-law function of the thermal-spike energy. This results in an overall decrease of the lattice thermal conductivity compared to that of the pristine glassy structure. Our simulations suggest that radiation-induced cascades can therefore reduce the lattice thermal conductivity of glassy \ce{Ge2Sb2Te5} by as much as 60\%. Given the remarkable recovery of the electronic structure of the glass after irradiation, as already demonstrated for this material in our previous work,~\cite{Konstantinou2018} ion irradiation can therefore be a potential strategy for improving the performance of phase-change memory and data-storage devices by reducing thermal cross-talk between memory cells. While the qualitative trends should stay the same, in the future, these effects should be explored in larger models which allow for better statistics and a wider range of higher thermal-spike energies, as well as in models of the crystalline phases of \ce{Ge2Sb2Te5}.\\

See \href{[URL will be inserted by publisher]}{Supplemental Material} for details about the thermal profiles established in the asymptotic regime in the irradiated glass, the calculation of the heat capacity in the simulated systems and the effect of the Langevin-thermostat damping time during irradiation on the calculation of the thermal conductivity.\\

F.C.M. acknowledges financial support from the UK Engineering and Physical Sciences Research Council (EPSRC) Centre for Doctoral Training in Computational Methods for Materials Science under grant EP/L015552/1, and resources provided by the ``Cambridge Service for Data Driven Discovery" (CSD3, http://csd3.cam.ac.uk) system operated by the University of Cambridge Research Computing Service (http://www.hpc.cam.ac.uk ) and funded by EPSRC Tier-2 capital grant EP/P020259/1. Via our membership of the UK's HEC Materials Chemistry Consortium, which is funded by EPSRC (EP/L000202, EP/R029431), this work used the ARCHER UK National Supercomputing Service (http://www.archer.ac.uk). K.K. acknowledges financial support from the EPSRC grant EP/N022009/1 (``Development and Application of Non-Equilibrium Doping in Amorphous Chalcogenides''), and also acknowledges the use of the High Performance Computing Facility (Grace@UCL), and associated support services, in the completion of this work.

\section*{References}
\bibliography{tc}%

%

\end{document}


\title{Supplementary Material: A first-principles study on the lattice thermal conductivity of irradiated glassy states of the \ce{Ge2Sb2Te5} phase-change memory material }

\author{Felix C. Mocanu}
\email[]{fcm29@cam.ac.uk}
\affiliation{Department of Chemistry, University of Cambridge, Lensfield Road, CB2 1EW, Cambridge, United Kingdom}

\author{Konstantinos Konstantinou}
\affiliation{Department of Chemistry, University of Cambridge, Lensfield Road, CB2 1EW, Cambridge, United Kingdom}

\author{Stephen R. Elliott}
\affiliation{Department of Chemistry, University of Cambridge, Lensfield Road, CB2 1EW, Cambridge, United Kingdom}

\maketitle

\clearpage

\section*{\label{Profile} Thermal Profiles}

\begin{figure}[!ht]
	\includegraphics[width=0.6\linewidth]{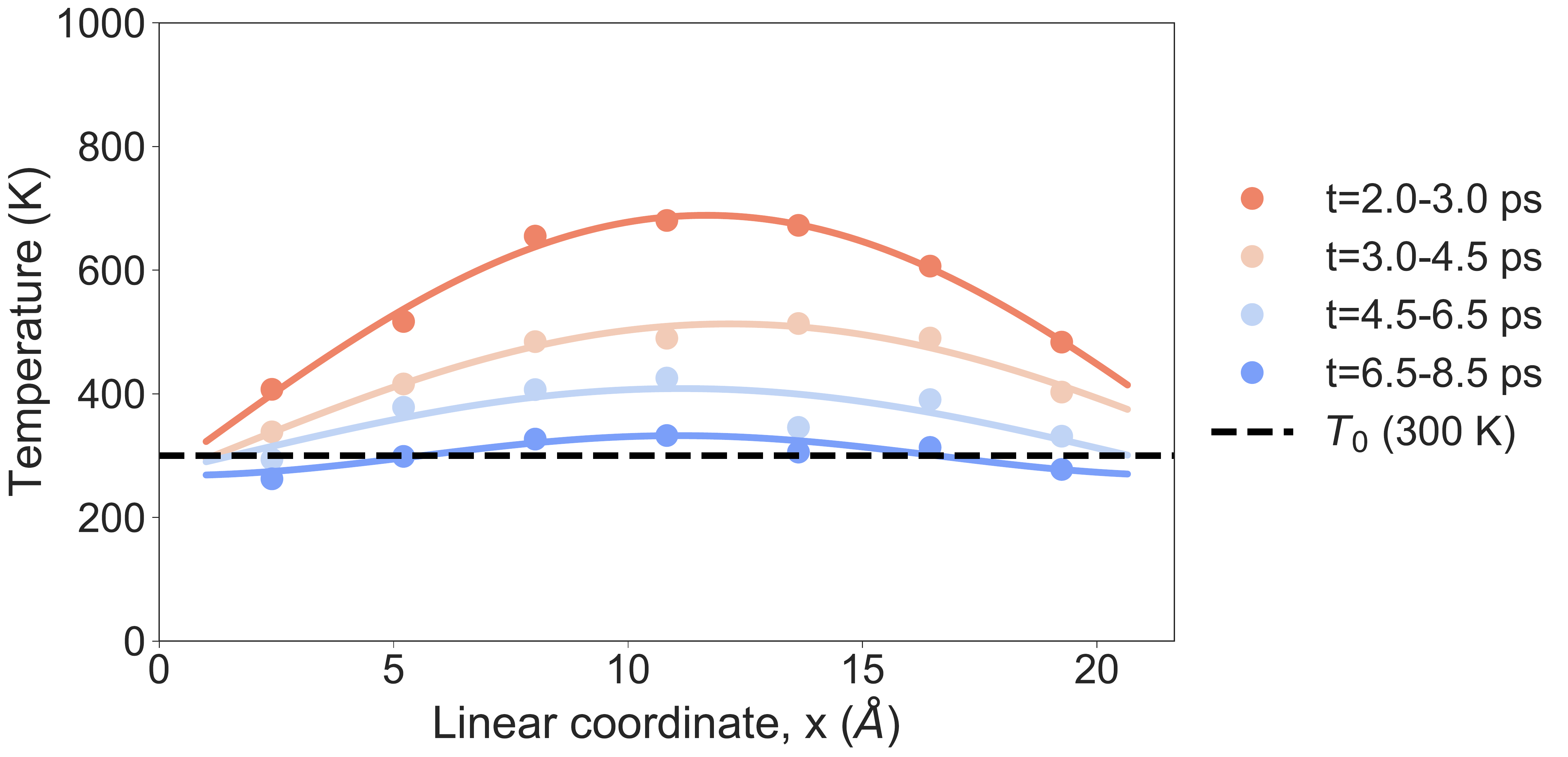}
	\includegraphics[width=0.6\linewidth]{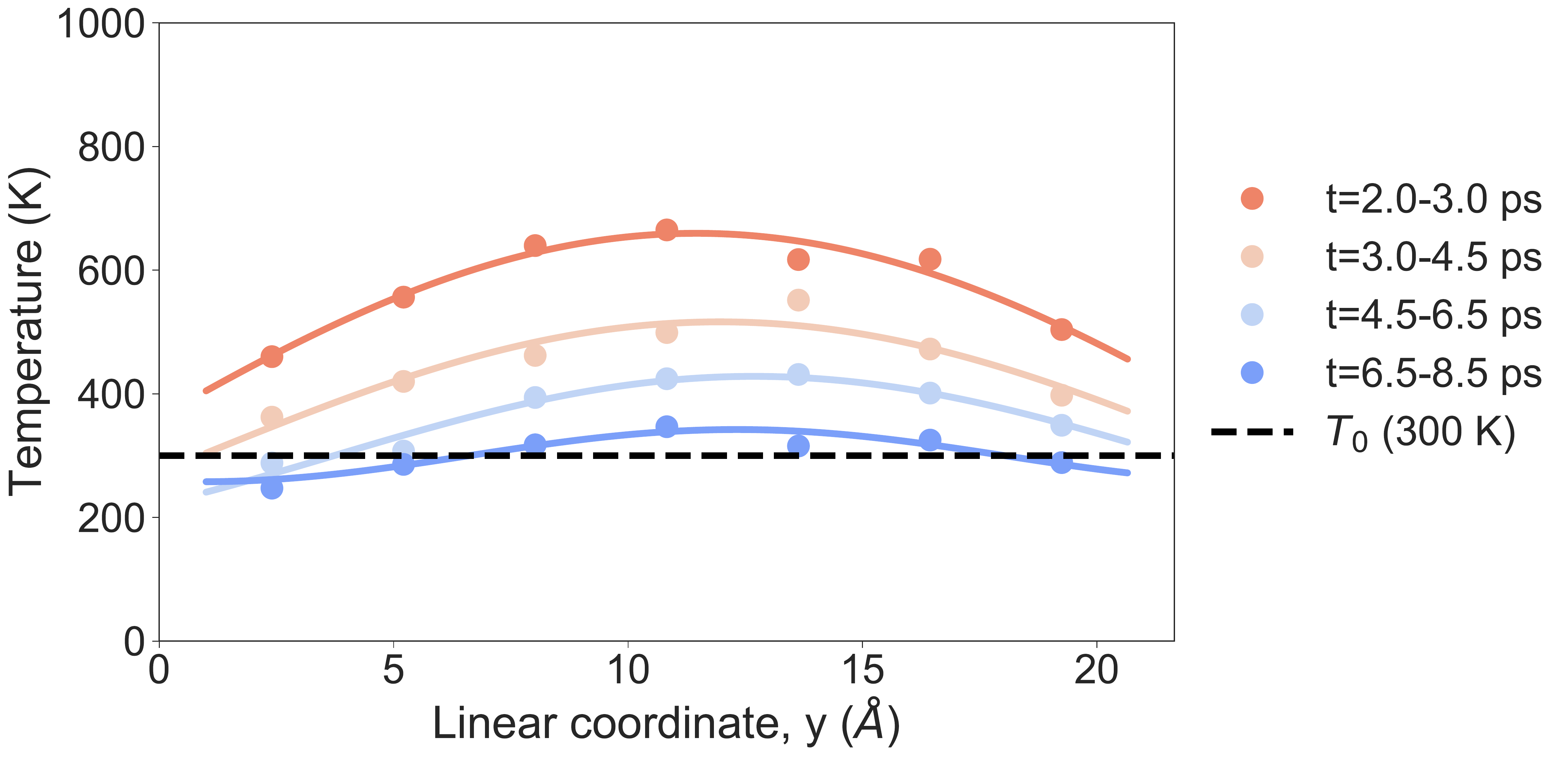}
	\includegraphics[width=0.6\linewidth]{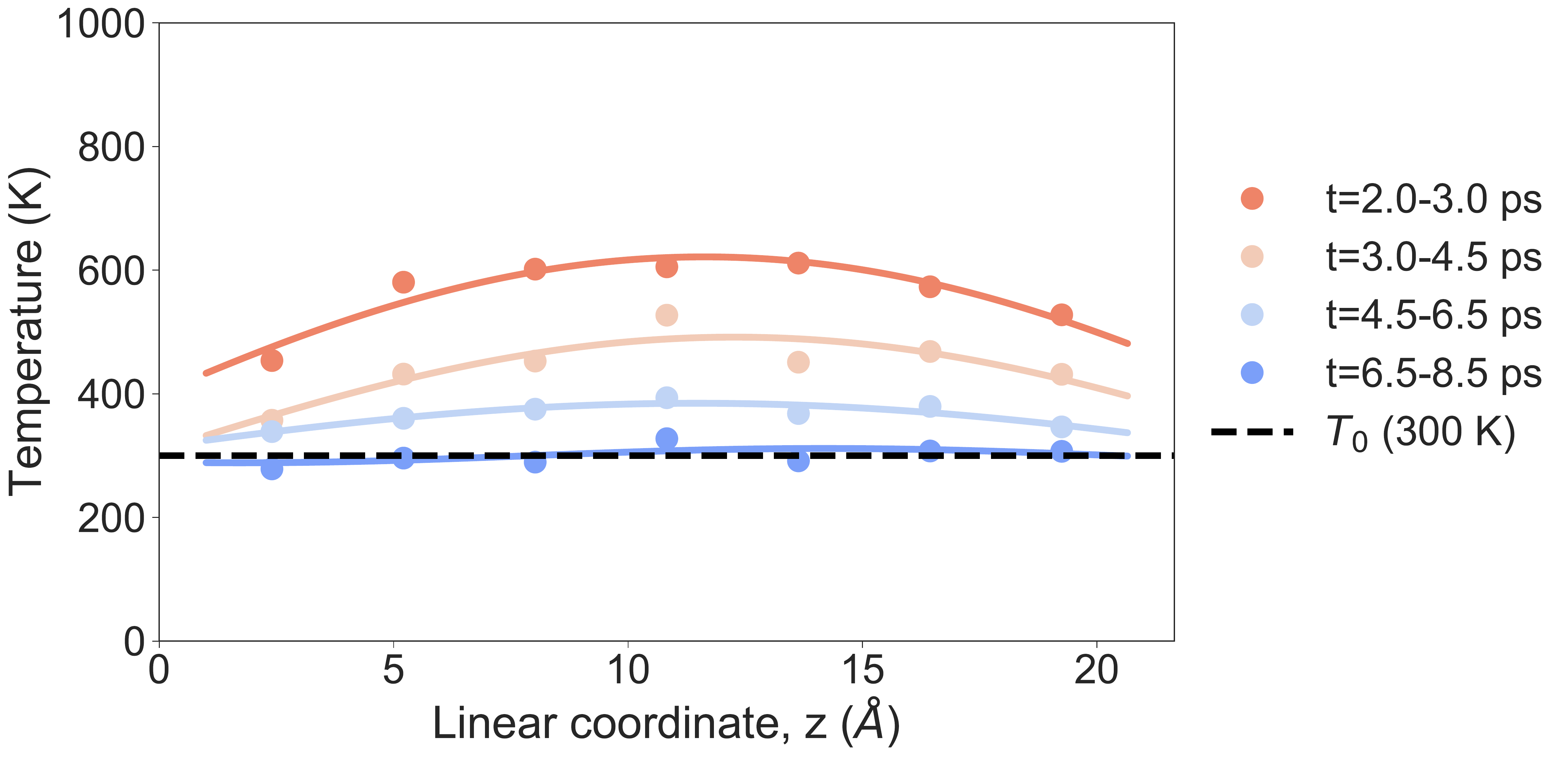}
	\caption{Thermal profiles, along different Cartesian directions, for the irradiated glass model with an initial thermal-spike energy of 150 eV, approaching equilibrium with the boundary layer. Dots represent kinetic temperature averages along a linear profile coloured from red to blue as the core cools down and approaches equilibrium with the boundary. The solid lines represent sinusoidal fits to the data.}
	\label{fig:profiles}
\end{figure}

Figure~\ref{fig:profiles} shows the sinusoidal thermal profiles established in the approach-to-equilibrium of the core with the thermostatted boudnary. The profiles were taken from the irradiated glass structure with an initial thermal-spike energy of $150$ eV over a linear profile in each cartesian direction and averaged over time intervals of $4000$ time steps each. As expected, the temperature is lower close to the boundary and higher in the centre of the cell. The profile deviates from an ideal sinusoidal curve that reaches $300$ K at the boundary along every direction. This is due to the small size of the model, the lack of uniformity of the thin boundaries and the artificial Kapitza resistance which results in a thermal jump at the interface. This discontinuity in temperature at the interface is related to the thermal gradient and is influenced by the Langevin-thermostat damping time, $\epsilon$, and the size of the simulation box, $L$.~\cite{Singh2015a} 

\clearpage

\section*{\label{thermostat} Thermostat effects}

 A proposed phenomenological model for the ``direct" non-eqilibrium molecular dynamics (NEMD) method of estimating the thermal conductivity relates the bulk thermal conductivity, $\kappa_{\infty}$, to the one obtained from a finite-sized simulation with a characteristic simulation length $L$, and using a particular thermostat damping time, $\epsilon$.~\cite{Singh2015a}
 
 \begin{equation}
 \label{eq:pheno}
 \dfrac{1}{\kappa} = \dfrac{1}{\kappa_{\infty}} + \dfrac{c\epsilon^p}{L}
 \end{equation}
 
 Since all the models simulated in this work have the same size (315-atom glass structures within a $L=21.651$ \AA{} cubic simulation box) the terms relating to the finite-size effects were merged into a single constant $c_L$. The simplified phenomenological model then reduces to:

\begin{equation}
\label{eq:thermostat}
\dfrac{1}{\kappa} = \dfrac{1}{\kappa_{\infty}} + c_L\epsilon^p
\end{equation}

The extrapolation due to the thermostat damping time cannot be taken as a substitute for estimating the finite-size effects. A larger model, with a proportionally thicker boundary may be needed to obtain the correct heat transport between the core and boundary regions in order to fully satisfy the assumptions of short-range interactions \cite{Kantorovich2008}. Given that the thermal-spike energies considered in this study are relatively low, in the range of $15-200$ eV, both the ionic and electronic systems have been treated adiabatically. The stochastic nature of the Langevin thermostat that was used for the boundary region, means that averaging over several runs may be required to converge the results in each case. 

\begin{figure}[!ht]
	\includegraphics[width=0.7\textwidth]{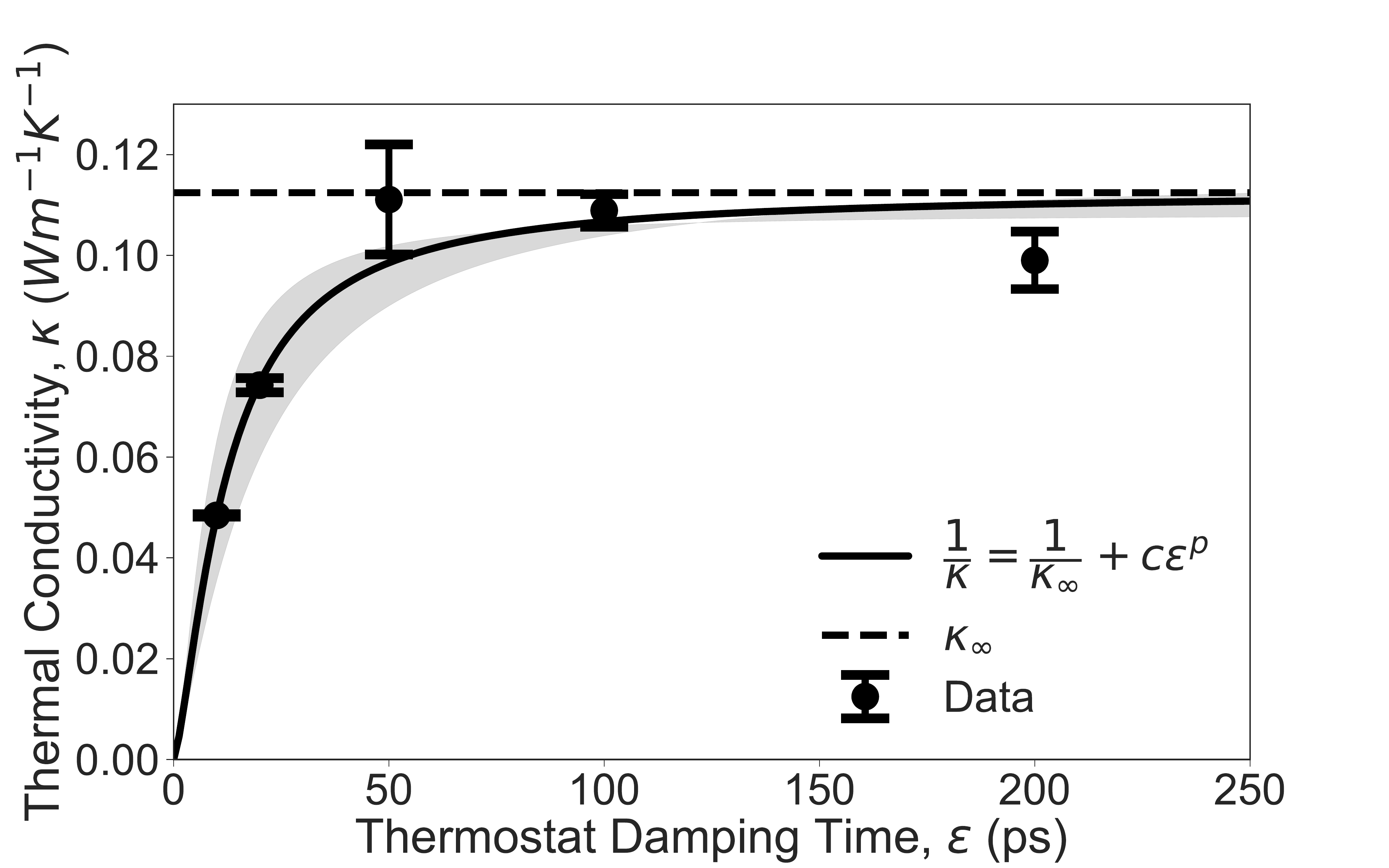}
	\caption{Thermal conductivity versus thermostat damping time. The extrapolated thermal conductivity is shown as a dashed line.}
	\label{fig:thermostat}
\end{figure}

One would expect that the magnitude of the temperature jump at the interface is reduced as the damping time increases and the thermostat couples more smoothly to the system, thus resulting in an increase in the estimated thermal conductivity. More importantly, the thermostatted boundary will approach NVE dynamics in the limit of infinite damping time. In order to evaluate these effects numerically, the thermal quench and subsequent coming into equilibrium with the boundary layer for the pristine glassy \ce{Ge2Sb2Te5} structure was repeated with several different thermostat damping times, in the interval $10-200$ fs. The data was used fit equation~\ref{eq:thermostat}, and to obtain an estimate for the lattice thermal conductivity which takes the thermostat effects into account. The obtained results can be seen in Figure \ref{fig:thermostat}, while the fitted parameters are $k_{\infty}=0.112 \pm 0.004$ $W K^{-1} m^{-1}$, $c_L=289 \pm 75$ m ps $^{1-p}$ K J$^{-1}$ and $p=-1.39 \pm 0.12$. For the irradiated glass samples, it is assumed that the fitting parameters $c_L$ and $p$ do not change. This allows us to get an estimate for the lattice thermal conductivity in the limit of infinitely gentle thermostatting for these glassy structures as well.  

The negative, as opposed to positive value of $p$, could be due to the simplification of the original model as well as due to the \ce{Ge2Sb2Te5} material itself, which has a much lower thermal conductivity when compared to the Lennard-Jones model of argon which was originally studied with this model.~\cite{Singh2015a} 

\section*{\label{Cv} Heat Capacity Estimation}

In a constant-volume, constant-temperature (NVT) molecular-dynamics simulation, the heat capacity can be extracted from the fluctuations in the total energy $E$, which are properly defined within the simulation and can be written as a function of volume, $V$, and  temperature, $T$, as:

\begin{align}
\delta E &= \left( \dfrac{\partial E}{\partial V} \right)_T \delta V + \left( \dfrac{\partial E}{\partial T} \right)_V \delta T = \left[ T \left( \dfrac{\partial P}{\partial T} \right)_V - P \right] \delta V + C_v \delta T
\end{align}

Assuming Gaussian statistics, the two sides of the equation can be squared and averaged to obtain the mean-square fluctuations in total energy:

\begin{equation}
\langle \delta E^2 \rangle = - \left[ T \left( \dfrac{\partial P}{\partial T} \right)_V - P \right]^2 T \left( \dfrac{\partial V}{\partial P} \right)_T + C_v T^2
\end{equation}
By taking into account that the fluctuations in volume are zero, $\delta V = 0$, due to the NVT molecular-dynamics simulation, then the heat capacity can be calculated from the expression:

\begin{align}
\label{eq:cv}
C_v = \dfrac{\langle \delta E^2\rangle}{k_B T^2}
\end{align}

\begin{figure}[!ht]
	\includegraphics[width=0.7\textwidth]{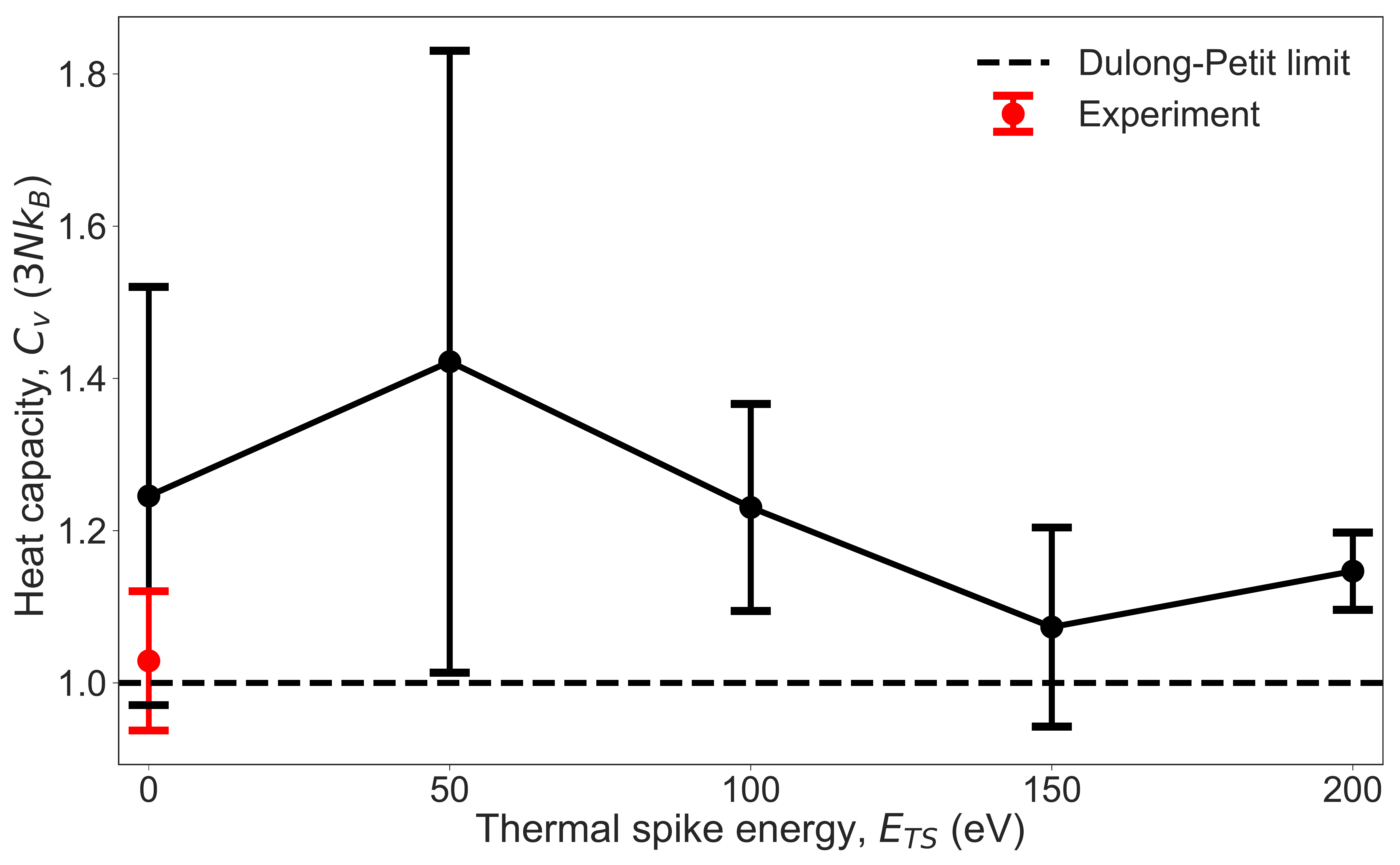}
	\caption{Heat capacities in units of $3Nk_B$ versus the energy of the thermal spike. The Dulong-Petit limit is shown as a dashed line.}
	\label{fig:cv}
\end{figure}

The sampling of the total-energy fluctuations, $\delta E$, is greatly improved through the use of a generalized Langevin equation (GLE), or colored-noise, thermostat.~\cite{Ceriotti2009, Ceriotti2010}
The optimal sampling parameters for the glassy material under study are obtained from the \url{www.gle4md.org} web form. A central frequency, $\omega_0$, of 3 THz was used for generating thermostat parameters, in accordance with the previously reported vibrational density of states of glassy \ce{Ge2Sb2Te5},~\cite{Caravati2009}  together with a ratio $\omega_{max} / \omega_{min}$ of $10^4$. It is noted that $\omega_{min}$ is a parameter of the thermostat and not the minimum vibrational frequency of the system.

Molecular-dynamics trajectories of $40$ ps were generated with a time-step of 1 fs for each glassy sample, totalling 200 ps of simulation time. For the irradiated glassy \ce{Ge2Sb2Te5} models, the starting configurations of these molecular dynamics trajectories are the optimized final structures obtained at the end of each non-equilibrium thermal-spike simulation when the core is in equilibrium with the boundary at 300 K. 

The first 8 ps from each 40 ps trajectory are discarded, in order to ensure that the new system, now with a single thermostatted region, has reached equilibrium. The heat capacity is then estimated, using equation \ref{eq:cv}, from the total energy fluctuations sampled from the remaining 32000 time steps. The calculated heat capacities for the pristine glassy phase and the irradiated glass models are shown in Figure \ref{fig:cv}. Error bars are calculated by splitting the trajectory into four equal intervals of 8 ps with no overlap, and calculating the mean and standard deviation. 

It can be seen that the heat capacities of the irradiated glassy structures, as calculated from energy fluctuations of NVT molecular-dynamics trajectories at 300 K, are above the Dulong-Petit limit,~\cite{Trachenko2010} and show a maximum before the thermal spike energies are large enough to fully melt the initial glassy structure. A similar maximum was observed for the thermal evolution of heat capacity in glassy solids, or viscous liquids (depending on the temperature).~\cite{Andritsos2013} While the increase in heat capacity corresponds to an increased anharmonicity of the self-irradiated models, in our case all models are (glassy) solids and the ultimate decrease in the heat capacity corresponds to a reduction of anharmonicity when the system is able to find a more harmonic glassy state. This happens because the natural quenches of the non-equilbrium cascades become longer with increasing thermal-spike energies and the system has enough time and kinetic energy to find a more harmonic minimum.

\section*{Summary of Results}

The results of all the calculations are summarised in table~\ref{tab:bulkthermal}.

\begin{table}[!ht]
	\caption{Calculated thermal relaxation times $\tau$, constant volume heat capacities $C_v$, lattice thermal conductivities $\kappa$, and $\kappa_{\infty}$, in the limit of infinitely gentle thermostatting of the boundary. Values are given for the pristine and irradiated glassy Ge$_2$Sb$_2$Te$_5$ models.}
	\label{tab:bulkthermal}
	\begin{ruledtabular}
		\begin{tabular}{l l l l l}
			$E_{TS}$ ($eV$) & $\tau$ (ps)& $C_v$ ($3Nk_B$) & $\kappa$ ($W K^{-1} m^{-1}$) & $\kappa_{\infty}$ ($W K^{-1} m^{-1}$) \\
			\hline
			$0$ (extrap.)    & $1.411 \pm 0.101$ & $1.246 \pm 0.275$ & $0.148 \pm 0.034$ & $0.160 \pm 0.040$\\
			$50$  & $2.226 \pm 0.011$ & $1.422 \pm 0.408$ & $0.107 \pm 0.031$ & $0.113 \pm 0.034$\\
			$100$ & $2.466 \pm 0.006$ & $1.231 \pm 0.136$ & $0.084 \pm 0.009$ & $0.087 \pm 0.010$\\
			$150$ & $2.745 \pm 0.004$& $1.073 \pm 0.131$ & $0.065 \pm 0.008$ & $0.067 \pm 0.009$\\
			$200$ & $2.776 \pm 0.003$& $1.147 \pm 0.051$ & $0.069 \pm 0.003$ & $0.071 \pm 0.003$\\
		\end{tabular}
	\end{ruledtabular}
\end{table}

\section*{References}

\bibliographystyle{aipnum4-1}
\bibliography{SI}%